\documentclass[12pt,a4paper]{article}
\usepackage[fleqn]{amsmath}
\usepackage{epsfig}
\usepackage{amssymb}
\usepackage{amsthm}

\numberwithin{equation}{section}

\newcommand{\bx}{{\boldsymbol x}}
\newcommand{\bX}{{\boldsymbol X}}

\begin{document}
\begin{flushright}
November, 2004
\end{flushright}

\bigskip

\begin{center}
{\large INTEGRABLE SYSTEMS AND DISCRETE GEOMETRY} 

\bigskip
\bigskip

{\bf Adam Doliwa$^{\dagger}$ and Paolo Maria Santini$^{\S}$} 

\bigskip

~${\dagger}$ Uniwersytet Warmi\'{n}sko--Mazurski w Olsztynie

\bigskip

${\S}$ Universit\`a di Roma 1 ``La Sapienza'' and 

\medskip

Istituto Nazionale di Fisica Nucleare, Sezione di Roma 1

\end{center}

\bigskip

\begin{flushleft}
{\it Send correspondence to:}

\bigskip

Adam Doliwa

Department of Mathematics and Computer Sciences

University of Warmia and Mazury in Olsztyn

ul.~\.{Z}o{\l}nierska 14 A, 10-561 Olsztyn, Poland 

Phone (+48 89) 524 60 71; FAX (+48 89) 534 02 50; 

E-mail {\tt doliwa@matman.uwm.edu.pl}

\end{flushleft}

\newpage

\section{Introduction}
\label{sec:intr}
Although the main subject of this article is
the connection between integrable discrete systems and geometry, we feel
obliged to begin with the differential part of the relation. 

\subsection{Classical differential geometry and integrable systems}
The oldest (1840) integrable nonlinear partial differential 
equation recorded in literature is the Lam\'e system 
\begin{gather}
\label{eq:Lame1}
\frac{\partial^2 H_i}{\partial u_j\partial u_k} - 
\frac{1}{H_j}\frac{\partial H_j}{\partial u_k}
\frac{\partial H_i}{\partial u_j} -
\frac{1}{H_k}\frac{\partial H_k}{\partial u_j}
\frac{\partial H_i}{\partial u_k}  = 0, 
\qquad  i,j,k \quad \text{distinct}
\\ 
\label{eq:Lame2}
\frac{\partial}{\partial u_k} 
\left(\frac{1}{H_k} \frac{\partial H_{j}}{\partial u_k} \right) +
\frac{\partial}{\partial u_j} 
\left(\frac{1}{H_j} \frac{\partial H_{k}}{\partial u_j} \right) +
\frac{1}{H_{i}^2}
\frac{\partial H_j}{\partial u_i} 
\frac{\partial H_k}{\partial u_i} =0,
\end{gather}
describing orthogonal coordinates 
in the three dimensional Euclidean
space~$\mathbb{E}^3$ (indices $i,j,k$ range from $1$ to $3$).
Already in 1869 it was found by Ribaucour that the \emph{nonlinear} 
Lam\'e system possesses a
discrete symmetry enabling to construct, in a \emph{linear} way, new
solutions of the system from the old ones.
He gave also a geometric interpretation of this symmetry  
in terms of certain spheres tangent to the coordinate surfaces of the 
triply orthogonal system. In 1918 Bianchi 
showed that the result of
superposition of the Ribaucour transformations is, in a certain sense, 
independ of the order of their composition.

Such properties of a nonlinear equation are hallmarks 
of its integrability, and
indeed, the Lam\'e system was solved using soliton techniques in 1997-98.
The above example illustrates the close connection between the modern theory
of integrable partial differential equations and the 
differential geometry of the turn of the XIXth and XXth centuries. 
A remarkable property of certain parametrized 
submanifolds (and then of the corresponding equations) 
studied that time is that they allow for transformations  
which exhibit the so called \emph{Bianchi permutability property}.
Such transformations called, depending on the
context, the Darboux, Calapso, Christoffel, 
Bianchi, B\"acklund, Laplace, Koenigs, Moutard, 
Combescure, L\'evy, Goursat, Ribaucour or the fundamental transformation of 
Jonas, can be geometrically described in terms of 
certain families of lines called line congruences. 
 
In the connection between integrable systems and differential geometry,
a distinguished role is played by the multidimensional 
conjugate nets, described by the Darboux system, which is just the first
part \eqref{eq:Lame1} of the Lam\'e system with indices ranging form $1$ to
$N\geq3$. On the level of 
integrable systems, this dominant role has the following
explanation: the Darboux system, together with equations
describing iso-conjugate deformations of the net, forms the 
multicomponent Kadomtsev--Petviashvilii (KP) hierarchy, which is
viewed as a master system of equations in soliton theory. In fact, in
appropriate variables, the whole
multicomponent KP hierarchy can be rewritten as an infinite system of the 
Darboux equations.
\subsection{Transition to the discrete domain}
The recent progress in studying discrete integrable systems showed that, in
many respects, they should be considered as more fundamental than their
differential counterparts. Consequently, the
natural problem of extending the geometric interpretation of
integrable partial differential equations to the discrete domain arose,
leading not only to the transition to the discrete domain of many results on
the connection between the differential geometry and integrable systems, 
but also -- and this seems to be even more important --
to the description of integrability in a very elementary and 
purely geometric way. 

On the level of integrable equations, the transition 'from
differential to discrete' often makes formulas more complicated and longer.
On the contrary, on the geometric level, in such a transition the properties of 
discrete submanifolds, relevant to their integrability, 
become simpler and more transparent. Indeed, the mathematics 
necessary to understand the basic ideas of the integrable discrete geometry
does not exceed the 'ruler and compass constructions', and many proofs can
be performed using elementary incidence geometry. 

We will concentrate our attention on
the multidimensional lattice made from planar quadrilaterals,  
which is the discrete analogue of a conjugate net. Together with the
discussion of its properties, which are the core of the geometric 
integrability, we briefly present
the analytic methods of construction of these lattices and we also 
describe some basic multidimensional
integrable reductions of them. 
Then we discuss integrable 
discrete surfaces; some of them have been found in the early period of the
'case by case' studies. We shall however try to
present them, from a unifying perspective, as reductions of
the quadrilateral lattice. 

\section{Multidimensional integrable lattices}
\label{sec:MQL}
\subsection{The quadrilateral lattice} 
An $N$ dimensional lattice $\bx : \mathbb{Z}^N \rightarrow \mathbb{R}^M$ 
is a 
lattice made from planar quadrilaterals, or a quadrilateral lattice 
(QL) in short,
if its elementary quadrilaterals 
$\{\bx ,T_i\bx , T_j\bx ,T_iT_j\bx \}$ are planar; i.e., iff       
the following system of discrete Laplace equations is satisfied
\begin{equation}  \label{eq:Laplace}
\Delta_i\Delta_j\bx=(T_{i} A_{ij})\Delta_i\bx+
(T_j A_{ji})\Delta_j\bx ,\quad i\ne j, \quad  i,j=1 ,\dots, N,
\end{equation}
where $A_{ij}:\mathbb{Z}^N\to\mathbb{R}$ are functions of the discrete variable;
here $T_i$ is the translation operator in the $i$th direction and 
$\Delta_i = T_i -1$ is the corresponding difference operator. For simplicity we
work here in the affine setting neglecting projective geometric aspects of 
the theory.
\subsubsection{The geometric integrability scheme}
\label{sec:geom-int}
\begin{figure}
\begin{center}
\epsffile{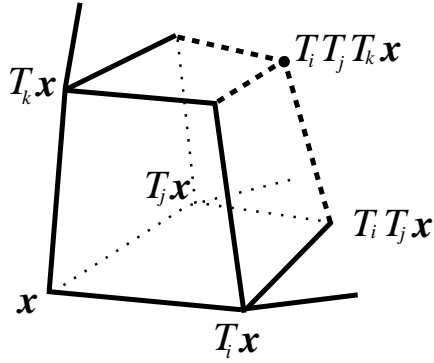}
\end{center}
\caption{The geometric integrability scheme}
\label{fig:TiTjTkx}
\end{figure}

In the case $N=2$ the definition \eqref{eq:Laplace} allows one to uniquely
construct, given two discrete curves intersecting in a common vertex and two
functions $A_{12},A_{21}:\mathbb{Z}^2\to\mathbb{R}$, a quarilateral surface. 
For $N>2$ the planarity constraints \eqref{eq:Laplace} are instead
compatible if and only if the geometric data $A_{ij}$ satisfy 
the nonlinear system
\begin{equation} \label{eq:MQL-A}
\Delta_k A_{ij} + (T_k A_{ij})A_{ik}=
 (T_jA_{jk})A_{ij} +(T_k A_{kj})A_{ik} ,
\qquad i, j, k \quad \text{distinct}.
\end{equation}
This constraint has very simple interpretation: in building the 
elementary cube (see Figure~\ref{fig:TiTjTkx}), 
the seven points $\bx$,
$T_i \bx$, $T_j \bx$, $T_k \bx$, $T_i T_j \bx$, $T_i T_k \bx$ and 
$T_j T_k \bx$ 
($i,j,k$ are distinct) determine the eighth point 
$T_i T_j T_k \bx$ as the unique intersection of three planes in the three 
dimensional space. 

The connection of this elementary geometric point of view with the 
classical theory of integrable systems is transparent: 
the planarity constraint corresponds to the set of \emph{linear} spectral 
problems \eqref{eq:Laplace} and the resulting QL 
is characterized by the nonlinear equations \eqref{eq:MQL-A}, arising as 
the compatibility conditions for such spectral 
problems. Since the QL equations \eqref{eq:MQL-A} are a master system in 
the theory of integrable equations, \emph{planarity} can be viewed as 
the \emph{elementary geometric root of integrability}. 
The idea that integrability be associated with the consistency of a 
geometric (and/or algebraic) property when increasing the dimensionality 
of the system is recurrent in the theory of integrable systems.

\subsubsection{Other forms of the Darboux system}
The $i\leftrightarrow j$ symmetry of the RHS of 
equations \eqref{eq:MQL-A}  
implies the existence of the potentials $H_i:\mathbb{Z}^N\to\mathbb{R}$ 
(the Lam\'e coefficients) such that
\begin{equation}
A_{ij}= \frac{\Delta_j H_i}{H_i}, \qquad i\ne j,
\end{equation}
and then equations \eqref{eq:MQL-A} 
take the form 
\begin{equation} \label{eq:MQL-H}
\Delta_k\Delta_jH_i - \left(T_j\frac{\Delta_kH_j}{H_j}\right)\Delta_jH_i -
\left(T_k \frac{\Delta_jH_k}{H_k}\right)\Delta_k H_i =0, 
\qquad i, j, k \quad \text{distinct},
\end{equation}
being the discrete version of the first part \eqref{eq:Lame1}
of the Lam\'e system.

The Lam\'e coefficients allow to define the suitably normalized tangent 
vectors $\bX_i:\mathbb{Z}^N\to\mathbb{R}^M$ by equations
\begin{equation} 
\label{def:H-Xi} \Delta_i \bx  = (T_i H_i) \bX_i , 
\end{equation}  
and the functions 
$Q_{ij}: \mathbb{Z}^N \to \mathbb{R}$, $i\ne j$, (the rotation coefficients) by equations
\begin{equation}
\label{eq:lin-H}
\Delta_iH_j =  (T_iH_i) Q_{ij}, \qquad i\ne j.
\end{equation}
Then equations \eqref{eq:Laplace} and \eqref{eq:MQL-H} can be 
rewritten in the first order form
\begin{gather} \label{eq:lin-X} 
\Delta_j\bX_i = (T_j Q_{ij})\bX_j , \qquad i\ne j, \\
\label{eq:MQL-Q}
\Delta_k Q_{ij} = (T_k Q_{ik}) Q_{kj} , \qquad i, j, k \quad \text{distinct}.
\end{gather}

The discrete Darboux system \eqref{eq:MQL-Q} implies the existence of
other potentials $\rho_i$ defined by the compatible equations
\begin{equation} \label{eq:rho-constr}
\frac{T_j\rho_i}{\rho_i} = 1 - (T_iQ_{ji})(T_jQ_{ij}) , \qquad
i\ne j  .
\end{equation}
The $i\leftrightarrow j$ symmetry of the RHS of 
equations \eqref{eq:rho-constr} implies 
existence of yet another potential $\tau:\mathbb{Z}^N\to \mathbb{R}$, 
\begin{equation}
\rho_i = \frac{T_i\tau}{\tau} ,
\end{equation}
which is called the $\tau$-function of the quadrilateral lattice. 
In terms of the $\tau$ function, and of the functions
\begin{equation}
\tau_{ij} = \tau Q_{ij} ,\qquad i\ne j,
\end{equation}
whose geometric interpretation will be given in 
Section~\ref{sec:Lapl-Hir}, the discrete Darboux equations take the 
following Hirota-type form 
\begin{gather} \label{eq:Hir-ij}
(T_i T_j \tau ) \tau = 
(T_i\tau) T_j\tau - (T_i\tau_{ji}) T_j\tau_{ij} , \qquad i\ne j,\\
\label{eq:Hir-ijk}
(T_k \tau_{ij}) \tau = 
(T_k\tau) \tau_{ij} + (T_k\tau_{ik}) \tau_{kj}  ,
\qquad i, j, k \quad \text{distinct}   .
\end{gather}

\subsection{Analytic methods}
We will show how one can construct large classes of 
solutions of the discrete Darboux equations and the corresponding
quadrilateral lattices using two
basic analytical methods of the soliton theory: the 
$\bar\partial$ dressing method and the algebro-geometric techniques.
\subsubsection{The $\bar\partial$ dressing method}
Consider the non-local $\bar\partial$-problem 
\begin{equation} \label{eq:db-nonl}
\bar\partial \chi(z) + 
(\hat{R}\chi)(z) = 
\bar\partial \nu(z), \qquad 
\lim_{|z|\to\infty} \left(\chi(z) -\nu(z)
\right) =0,
\end{equation}
where $\bar\partial = \partial/\partial\bar{z}$, 
$\hat{R}$ is the integral operator
\begin{equation*}
(\hat{R}\chi)(z) = \int_\mathbb{C} 
R(z,z')\chi(z') \: 
dz'\wedge d\bar{z}',
\end{equation*}
and $\nu(z)$ is a given rational function of $z$.

Let $Q_i^\pm\in\mathbb{C}$, $i=1,\dots ,N$ be pairs of distinct
points of the complex plane, which define the 
dependence of the kernel $R$ on the discrete variable
$n\in\mathbb{Z}^N$
\begin{equation*} \label{eq:db-nonl-evol}
R(z,z';n) = 
\prod_{i=1}^N \left(\frac{z-Q_i^+}{z-Q_i^-} \right)^{n_i}
R_0(z,z') 
\prod_{i=1}^N \left(\frac{z'-Q_i^-}{z'-Q_i^+} \right)^{n_i}.
\end{equation*}
We consider only kernels $R_0(z,z')$ such that the non-local 
$\bar\partial$-problem is uniquely solvable. 
If $\chi(z;n)$ is the unique solution with the canonical normalization 
$\nu=1$, then the function
\begin{equation*}
\psi(z;n) = \chi(z;n)\prod_{i=1}^N 
\left(\frac{z-Q_i^-}{z-Q_i^+} \right)^{n_i}
\end{equation*}
satisfies the system of the Laplace equations \eqref{eq:Laplace}
with the Lam\'e
coefficients given by
\begin{equation*}
H_i(n) = \lim_{z\to Q_i^+} 
\left(\left(\frac{z-Q_i^+}{z-Q_i^-} \right)^{n_i} \psi(z;n) \right).
\end{equation*}
By construction, the system of such Laplace equations is compatible, therefore 
the Lam\'e coefficients satisfy equations \eqref{eq:MQL-H}. To various
$n$-independent measures $d\mu_a$ on $\mathbb{C}$ there correspond coordinates
\begin{equation*}
x^a(n)=\int_\mathbb{C} \psi(z;n) d\mu_a(z),
\end{equation*} 
of a quadrilateral lattice $\bx$, having $H_i(n)$ as the Lam\'e coefficients.
To have real lattices, the kernel $R_0$, the points $Q_i^\pm$ and the
measures $d\mu_a$ should satisfy certain additional conditions.

One can find a similar interpretation of the normalized tangent vectors
$\bX_i$ and of the rotation coefficients $Q_{ij}$. 
If $\chi_i(z;n)$ are the unique solutions of the non-local 
$\bar\partial$-problem \eqref{eq:db-nonl} with the normalizations
\begin{equation*}
\nu_i(z;n)=\left( \frac{Q_i^+ - Q_i^-}{z-Q_i^+} \right) 
\prod_{k=1, k\ne i}^{N} 
\left( \frac{Q_i^+ - Q_k^+}{Q_i^+ -Q_k^-} \right) ^{n_k},
\end{equation*}
then the functions $\psi_i(z;n)$, defined by
\begin{equation*}
\psi_i(z;n) = \prod_{k=1}^N 
\left( \frac{z - Q_k^-}{z-Q_k^+} \right) ^{n_k} \chi_i(z;n) ,
\end{equation*}
satisfy the direct
analogue of the linear problem \eqref{eq:lin-X}
\begin{equation} \label{eq:lin-X-db}
\Delta_j \psi_i(z;n) = (T_j Q_{ij}(n)) \psi_j(z;n), 
\qquad i\ne j ,
\end{equation}
where
\begin{equation*}
Q_{ij}(n) = 
\lim_{z\to Q_j^+} 
\left(\left(\frac{z-Q_j^+}{z-Q_j^-} \right)^{n_j} \psi_i(z;n) 
\right).
\end{equation*}
Again, by construction, equations \eqref{eq:lin-X-db} are compatible 
and the functions $Q_{ij}(n)$ satisfy the discrete Darboux equations
\eqref{eq:MQL-Q}.
The functions
\begin{equation*}
X_i^a(n)=\int_\mathbb{C} \psi_i(z;n) d\mu_a(z),
\end{equation*} 
are coordinates of the normalized tangent vectors $\bX_i$
of the quadrilateral lattice $\bx$ constructed above.

\subsubsection{The algebro-geometric techniques}

Given a compact Riemann surface $\mathcal{R}$ of genus $g$, consider 
a non-special divisor $D=\sum_{\alpha=1}^g P_\alpha$. Choose 
$N$ pairs of points $Q_i^\pm\in \mathcal{R}$ and the normalization
point $Q_\infty$. Given $n\in\mathbb{Z}^N$, there exists a unique
Baker--Akhiezer function 
$\psi(n)$, defined
as a meromorphic function on $\mathcal{R}$, with the following 
analytical properties: (i)~as a function of $P\in \mathcal{R}\setminus 
        \cup_{i=1}^N Q_i^\pm$, $\psi(n)$ 
        may have as singularities only simple poles in the points of the 
	divisor $D$;
(ii)~in the points $Q_i^\pm$ function $\psi(n)$ has poles
        of the order $\pm n_i$; 
(iii)~in the point $Q_\infty$  function $\psi(n)$ is normalized to $1$.

When $z_i^\pm(P)$ is a local coordinate on $\mathcal{R}
$ centered at $Q_i^\pm$, then 
condition (ii) implies that the function
$\psi(n)$ in a neigbourhood of the point $Q_i^\pm$ is of the form
\begin{equation}
\psi(P;n) = \left(z_i^\pm(P)\right)^{\mp n_i}\left( \sum_{s=0}^\infty
\xi_{s,\pm}^i(n)\left(z_i^\pm(P)\right)^{s} \right).
\end{equation}  
The Baker--Akhiezer function, 
as a function of the discrete variable $n\in\mathbb{Z}^N$, satisfies
the system of Laplace equations \eqref{eq:Laplace}
with the Lam\'e coefficients $H_i(n)=	\xi_{0,+}^i(n)$.

Again, by construction, the 
Lam\'e coefficients satisfy equations \eqref{eq:MQL-H}. To various
$n$-independent measures $d\mu_a$ on $\mathcal{R}$ there correspond 
coordinates
\begin{equation*}
x^a(n)=\int_\mathcal{R} \psi(P;n) d\mu_a(P),
\end{equation*} 
of a quadrilateral lattice $\bx$. 

We present the expression of the Baker--Akhiezer
function and of the $\tau$-function of the quadrilateral lattice in terms of
the Riemann theta functions. 
Let us choose on $\mathcal{R}$ the canonical basis
of cycles $\{a_1,\dots, a_g, b_1, \dots , b_g\}$ and the dual basis
$\{ \omega_1,\dots ,\omega_g\}$ of holomorphic 
differentials on $\mathcal{R}$, i.e.,
$
\oint_{a_j}\omega_k =  \delta_{jk}.
$
Then the matrix $B$ of $b$-periods defined as
$
B_{jk}=\oint_{b_j}\omega_k,
$
is symmetric and has positively defined
imaginary part.
Denote by $\omega_{PQ}$ the unique 
differential holomorphic in $\mathcal{R}\setminus\{P,Q\}$ 
with poles of the first
order in $P$, $Q$ and residues, correspondingly, $1$ and $-1$, which is
normalized by conditions  
$
\oint_{a_j}\omega_{PQ} =  0.
$
The Riemann function $\theta(z;B)$, $z\in\mathbb{C}^g$, is defined by
its Fourier expansion
\begin{equation*}
\theta(z;B) = \sum_{m\in{\mathbb{Z}^g}} \exp \left\{ \pi i \langle m,B m
\rangle + 2 \pi i\langle  m, z \rangle \right\},
\end{equation*}
where $\langle \cdot , \cdot \rangle$ denotes the standard bilinear
form in $\mathbb{C}^g$.
Finally, the Abel map $A$ is given by
$A(P) = \left ( \int_{P_0}^P \omega_1, \dots, \int_{P_0}^P \omega_g \right)$,
where $P_0 \in \mathcal{R}$, and  the  Riemann 
constants vector $K$ is given by
\begin{equation*}
K_j = \frac{1+B_{jj}}{2} -\sum_{k\ne j} \left( \oint_{a_k} \omega_k(P)
A_j(P) \omega_j   \right), \qquad j=1,\dots, g.
\end{equation*} 
The explicit
form of the vacuum Baker--Akhiezer function $\psi$ can be written down 
with the help of the theta functions as follows
\begin{multline*}
\psi(n,P) = \frac{\theta \left( A(P) + 
\sum_{k=1}^N n_k \left( A(Q_k^-) - A(Q_k^+)  \right) + Z   \right)}
{\theta \left( A(Q_\infty)+
\sum_{k=1}^N n_k \left( A(Q_k^-) - A(Q_k^+)   \right) + Z   \right)} 
\times \\ 
\times \frac{\theta\left( A(Q_\infty)+Z   \right)}
{\theta\left( A(P) + Z   \right)}
\exp \left( \sum_{k=1}^N n_k \int_{Q_\infty}^P \omega_{Q_k^- Q_k^+} \right), 
\qquad \qquad 
\end{multline*}
where $Z = - \sum_{j=1}^g A(P_j) - K$.

Denote by $r_{kj}^\pm$ and $s_{kj}^\pm$
the constants in the
decomposition of the Abelian integrals near the point $Q_j^\pm$
\begin{align*}
\int_{P_0}^P \omega_{Q_k^- Q_k^+} & \stackrel{P\to Q_j^\pm}{=} \mp
\delta_{kj} \log z_j^\pm(P) + r_{kj}^\pm + O(z_j^\pm(P) ), \\
\int_{P_0}^P \omega_{Q_\infty Q_k^+} & \stackrel{P\to Q_j^\pm}{=} -
\delta_{kj} \delta_{+\pm} \log z_j^\pm(P) + s_{kj}^\pm + 
O(z_j^\pm(P) ). 
\end{align*}
Then the expression of the $\tau$--function of the quadrilateral lattice 
within
the subclass of algebro-geometric solutions reads
\begin{equation*}
\tau(n) = \theta\left(  
\sum_{k=1}^N n_k\left( A(Q_k^-) - A(Q_k^+)  \right) +A(Q_\infty)
+ Z   \right) \prod_{k,j=1}^N \lambda_{kj}^{n_k n_j}
\prod_{k=1}^N \mu_{k}^{n_k}, 
\end{equation*}
where
\begin{equation*}
\lambda_{kj} = \exp \left( \frac{r_{kj}^- - r_{kj}^+}{2}
\right)=\lambda_{j k} ,
\qquad \mu_{k} = \frac{1}{\lambda_{kk}}
\frac{\theta\left(A(Q_k^+)+ Z   \right)} 
{\theta\left(A(Q_k^-)+ Z   \right)}
\exp \left( s_{kk}^- - s_{kk}^+ \right).
\end{equation*}

Finally, we remark that the geometric integrability scheme and the
algebro-geometric methods work also in the 
finite fields setting, giving solutions of the corresponding
integrable cellular automata. 

\subsection{The Darboux-type transformations}
We present the basic ideas and results of the theory of the Darboux type
transformations of the multidimensional quadrilateral 
lattice. 

\subsubsection{Line congruences and the fundamental transformation}
\label{sec:congr-fund}
To define the transformations we need to define first 
$N$-dimensional \emph{line congruences}
(or, simply, congruences), which are familes of lines in $\mathbb{R}^M$ labelled by
points of $\mathbb{Z}^N$ with the property that any two neighbouring lines
$\mathfrak{l}$ and $T_i \mathfrak{l}$, $i=1,...,N$, are coplanar 
and therefore (eventually in the projective extension $\mathbb{P}^M$ of 
$\mathbb{R}^M$) intersect.

\begin{figure}
\begin{center}
\epsffile{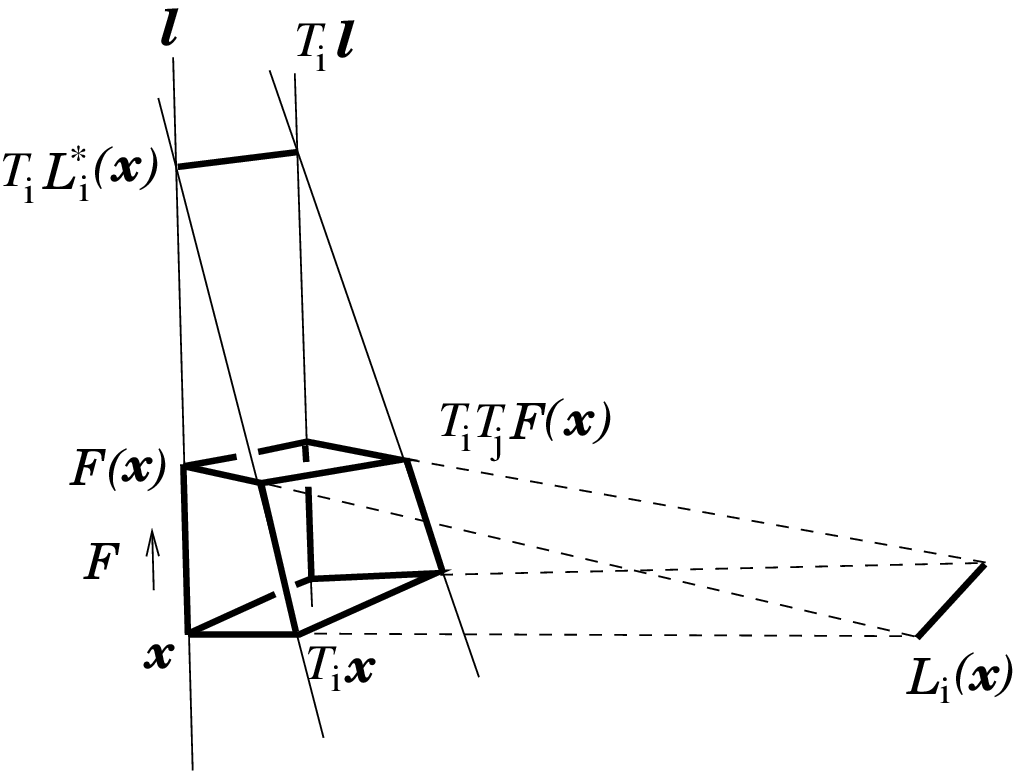}
\end{center}
\caption{The fundamental transformation as 
the binary transformation}
\label{fig:binary}
\end{figure}

The quadrilateral lattice $\mathcal{F}(\bx)$ is a \emph{fundamental
transform} of the quadrilateral lattice $\bx$ if the lines connecting the
corresponding points of the lattices form a congruence. The superposition of
a number of fundamental transformations can be compactly formulated in the
vectorial fundamental transformation.
The data of the vectorial fundamental
transformation are: (i)~the solution $\boldsymbol{Y}_i:\mathbb{Z}^N\to\mathbb{V}$, 
$\mathbb{V}$ being a
linear space, of the linear system \eqref{eq:lin-X}; (ii)~the solution 
$\boldsymbol{Y}^*_i:\mathbb{Z}^N\to\mathbb{V}^*$, $\mathbb{V}^*$ being the dual of 
$\mathbb{V}$, of the linear system \eqref{eq:lin-H}. These allow to
construct the linear operator valued potential 
$\boldsymbol{\Omega}(\boldsymbol{Y},\boldsymbol{Y}^*):
\mathbb{Z}^N\to L(\mathbb{V})$,
defined by the following analogue of equation \eqref{def:H-Xi}
\begin{equation}
\Delta_i\boldsymbol{\Omega}(\boldsymbol{Y},\boldsymbol{Y}^*) = 
\boldsymbol{Y}_i \otimes (T_i \boldsymbol{Y}^*_i), 
\qquad i = 1,\dots , N;
\end{equation} 
similarly, one defines 
$\boldsymbol{\Omega}(\boldsymbol{X},\boldsymbol{Y}^*):
\mathbb{Z}^N\to L(\mathbb{V},\mathbb{R}^M)$ and 
$\boldsymbol{\Omega}(\boldsymbol{Y},H):
\mathbb{Z}^N\to \mathbb{V}$.
The transforms of the lattice $\bx$ and other related functions
are given by
\begin{align*}
\boldsymbol{\mathcal{F}}(\bx) & = \bx - 
\boldsymbol{\Omega}(\boldsymbol{X},\boldsymbol{Y}^*)
\boldsymbol{\Omega}(\boldsymbol{Y},\boldsymbol{Y}^*)^{-1}
\boldsymbol{\Omega}(\boldsymbol{Y},H), \\
\boldsymbol{\mathcal{F}}(H_i) & = H_i -
\boldsymbol{Y}^*_i
\boldsymbol{\Omega}(\boldsymbol{Y},\boldsymbol{Y}^*)^{-1}
\boldsymbol{\Omega}(\boldsymbol{Y},H), \qquad i=1,\dots, N,\\
\boldsymbol{\mathcal{F}}(\bX_i) & = \bX_i - 
\boldsymbol{\Omega}(\boldsymbol{X},\boldsymbol{Y}^*)
\boldsymbol{\Omega}(\boldsymbol{Y},\boldsymbol{Y}^*)^{-1}
\boldsymbol{Y}_i, \qquad i=1,\dots, N,\\
\boldsymbol{\mathcal{F}}(Q_{ij}) & = Q_{ij} - 
\boldsymbol{Y}^*_j
\boldsymbol{\Omega}(\boldsymbol{Y},\boldsymbol{Y}^*)^{-1}
\boldsymbol{Y}_i, \qquad i,j=1,\dots, N, \quad i\ne j,\\
\boldsymbol{\mathcal{F}}(\rho_i) & = \rho_i\left( 
1 + (T_i\boldsymbol{Y}^*_i) 
\boldsymbol{\Omega}(\boldsymbol{Y},\boldsymbol{Y}^*)
\boldsymbol{Y}_i \right), \qquad i=1,\dots, N,\\
\boldsymbol{\mathcal{F}}(\tau) & = \tau \det
\boldsymbol{\Omega}(\boldsymbol{Y},\boldsymbol{Y}^*).
\end{align*}
Notice that, by the coplanarity of any two neighbouring lines of the 
congruence,
also the quadrilaterals  
$\{ \bx, T_i\bx,\mathcal{F}(\bx), \mathcal{F}(T_i\bx) \}$ are planar (see
Figure~\ref{fig:binary}).
Then the construction of the transformed
lattice mimics the geometric integrability scheme. 
In consequence, any quadrilateral 
$\{ \bx,\mathcal{F}_1(\bx), \mathcal{F}_2(\bx),
\mathcal{F}_1\left(\mathcal{F}_2(\bx)\right)=
\mathcal{F}_2\left(\mathcal{F}_1(\bx)\right)\}$ is planar as well. Therefore,
on the discrete level, there is no difference between the 
lattice coordinate directions and the fundamental transformation directions.
The distinction becomes visible in the limit from the quadrilateral lattice
to the conjugate net. Therefore 
the vectorial description of the superposition of the fundamental 
transformations not only implies their permutability but also provides the
explanation of the validity of the
practical rule of "integrable
discretization by Darboux transformations".

\subsubsection{The L\'evy and Combescure transformations}
\label{sec:fund-tr-red}
It is easy to see that the family $\mathfrak{t}_i$ of lines passing through
the points $\bx$ and $T_i\bx$ of a quadrilateral lattice forms a congruence,
called the $i$-th \emph{tangent congruence} of the lattice. When the
congruence of the transformation is the $i$-th tangent congruence of 
the lattice $\bx$, then the corresponding reduction of the fundamental
transformation is called the \emph{L\'evy transformation} $\mathcal{L}_i$. 

It turns out that, for a generic congruence $\mathfrak{l}$, the lattice made
from intersection points of the lines $\mathfrak{l}$ and 
$T_i^{-1}\mathfrak{l}$ is a quadrilateral lattice, called the $i$-th 
\emph{focal latice} of the congruence. When the fundamental transform
of the lattice $\bx$ is the $i$-th focal lattice of the transformation
congruence, then the corresponding reduction of the fundamental
transformation is called the \emph{adjoint L\'evy transformation} 
$\mathcal{L}_i^*$. 

Both L\'evy transformations use only a half of the fundamental
transformation data, and the corresponding reduction formulas (in the scalar
case) for the lattice points read as follows
\begin{align*}
\mathcal{L}_i(\bx) & = \bx -
\bX_i \, (Y_i)^{-1}\boldsymbol{\Omega}(Y,H),\\ 
\mathcal{L}^*_i(\bx) & = \bx - 
\boldsymbol{\Omega}(\bX,Y^*)\,(Y^*_i)^{-1}H_i.
\end{align*}
Notice that the composition of the L\'evy and the
adjoint L\'evy transformations gives (see Figure~\ref{fig:binary}) 
the fundamental
transformation, also called, for this reason, the binary transformation. 

Another reduction of the fundamental
transformation, important from a technical point of view, is 
the \emph{Combescure transformation}, in which the tangent lines of the 
transformed lattice
$\mathcal{C}(\bx)$ are parallel to those of
the lattice $\bx$. The transformation formula reads
\begin{equation*}
\mathcal{C}(\bx)  = \bx - 
\boldsymbol{\Omega}(\boldsymbol{X},{Y}^*),
\end{equation*} 
where only the solution $Y^*$ of the adjoint linear system~\eqref{eq:lin-H},
necessary to build the transformation congruence, is needed.
\subsubsection{The Laplace transformations and the geometric meaning 
of the Hirota equation}
\label{sec:Lapl-Hir}
\begin{figure}
\begin{center}
\epsffile{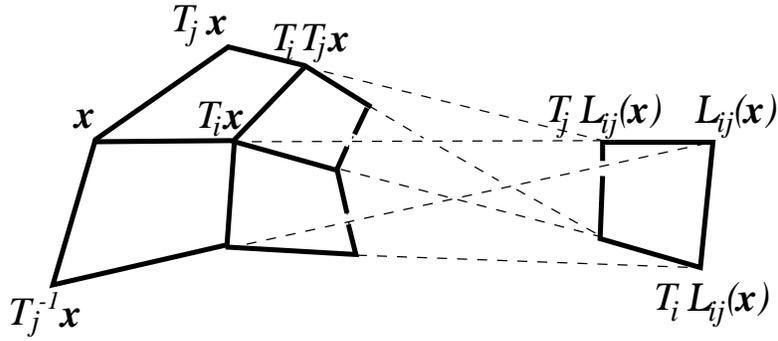}
\end{center}
\caption{The Laplace transformation $\mathcal{L}_{ij}$}
\label{fig:Lijx}
\end{figure}

The Laplace transform $\mathcal{L}_{ij}(\bx)$, $i\ne j$, of the quadrilateral
lattice $\bx$ is the $j$-th focal lattice of its $i$-th tangent congruence (see
Figure~\ref{fig:Lijx}).
It is uniquely determined once the lattice $\bx$ is given. The transformation
formulas of the lattice points and of the $\tau$-function read as follows
\begin{align}
\mathcal{L}_{ij}(\bx) & = \bx - \frac{1}{A_{ji}}\Delta_i\bx , \\
\mathcal{L}_{ij}(\tau) & = \tau_{ij} = \tau Q_{ij}. \label{eq:Lij-tau}
\end{align}
The superpositions of Laplace transformations satisfy the following
identities
\begin{equation*} 
\mathcal{L}_{ij} \circ \mathcal{L}_{ji} =  \text{id} , \qquad
\mathcal{L}_{jk} \circ \mathcal{L}_{ij} =  \mathcal{L}_{ik}, \qquad 
\mathcal{L}_{ki} \circ \mathcal{L}_{ij} =  \mathcal{L}_{kj},
\end{equation*}
which allow to identify them with the the Schlesinger transformations
of the monodromy theory.

In the simplest case $N=2$ 
one obtains the so called Laplace sequence of two dimensional
quadrilateral lattices
\begin{equation*}
\bx_{\ell} = \mathcal{L}_{12}^\ell(\bx), \qquad 
\tau_{\ell} = \mathcal{L}_{12}^\ell(\tau), \qquad 
\mathcal{L}_{12}^{-1} =\mathcal{L}_{21}, \quad \ell \in \mathbb{Z}.
\end{equation*}
Equations \eqref{eq:Hir-ij} and \eqref{eq:Lij-tau} imply that the
$\tau$-functions of the Laplace sequence satisfy the celebrated
Hirota equation (the fully discrete Toda system)
\begin{equation*} \label{eq:Htau}
\tau_{\ell} T_1T_2\tau_{\ell} = (T_1\tau_{\ell})(T_2\tau_{\ell}) -
(T_1 \tau_{\ell-1})(T_2\tau_{\ell+1}) .
\end{equation*}
\subsection{Distinguished integrable reductions}
We will present here basic reductions of the multidimensional quadrilateral
lattice. The geometric criterion for their integrability is the
compatibility with the geometric integrability scheme. 

\subsubsection{The circular lattices and the Ribaucour congruences}
Quadrilateral lattices $\mathbb{Z}^N\to\mathbb{E}^M$ for which each quadrilateral is
inscribed in a circle are called \emph{circular} lattices. 
They are the integrable
discrete analogues of submanifolds parametrized by curvature 
coordinates (for example, the orthogonal coordinate systems 
described by the Lam\'e
equations~\eqref{eq:Lame1}-\eqref{eq:Lame2}). 

\begin{figure}
\begin{center}
\epsffile{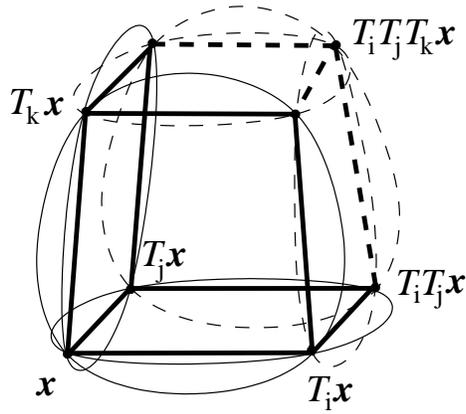}
\end{center}
\caption{The geometric integrability of circular lattices}
\label{fig:circ-red}
\end{figure}

The integrability of circular lattices is the consequence of the fact that,
if the three "initial" quadrilaterals 
$\{ \bx, T_i\bx, T_j\bx, T_i T_j \bx \}$,
$\{ \bx, T_i\bx, T_k\bx, T_i T_k \bx \}$,
$\{ \bx, T_j\bx, T_k\bx, T_j T_k \bx \}$, are circular, then also the three
new quadrilaterals constructed by adding the vertex $T_i T_j T_k \bx$, are 
circular as well (see Figure~\ref{fig:circ-red}). 
In fact, all the eight vertices belong to a sphere, and, in
consequence,
all the vertices of any $K$~dimensional, $K=2,\dots,N$, elementary cell belong 
to a $K-1$~dimensional sphere. 

There are various equivalent algebraic descriptions of the circular
lattices:\\
1. the normalized tangent vectors $\bX_i$ satisfy the constraint
\begin{equation*}
\bX_i \cdot T_i\bX_j + \bX_j \cdot T_j\bX_i = 0, \qquad i\ne j;
\end{equation*} 
2. the scalar function $\bx\cdot\bx:\mathbb{Z}^N\to\mathbb{R}$ satisfies the 
Laplace equations~\eqref{eq:Laplace} of the lattice $\bx$;\\
3. the functions $X^\circ_i=(\bx + T_i\bx)\cdot\bX_i:\mathbb{Z}^N\to\mathbb{R}$ satisfy the same
linear
system~\eqref{eq:lin-X} as the normalized tangent vectors $\bX_i$;\\
4. the functions $\bX_i\cdot\bX_i:\mathbb{Z}^N\to\mathbb{R}$ satisfy equations \eqref{eq:rho-constr}
and thus can serve as the potentials $\rho_i$.
 
The Ribaucour transformation $\mathcal{R}$
is the restriction of the
fundamental transformation to the class of circular lattices such that
also the "side" quadrilaterals 
$\{\bx, T_i\bx, \mathcal{R}(\bx), \mathcal{R}(T_i\bx)\}$ are circular. Again there is no geometric difference
between the lattice directions and the Ribaucour transformation direction.
Moreover, the quadrilaterals 
$\{ \bx, \mathcal{R}_1(\bx), \mathcal{R}_2(\bx), 
\mathcal{R}_1\left(\mathcal{R}_2(\bx)\right)=
\mathcal{R}_2\left(\mathcal{R}_1(\bx)\right) \}$ are circular as well.
In consequence, the vertices of the elementary $K$-cells, $K=2,\dots ,N$, of the
circular lattice and the corresponding vertices
of its Ribaucour transform are contained in a $K$~dimensional sphere.
Finally, for $K=N$, one obtains a special
$\mathbb{Z}^N$ family of $N$-dimensional spheres, called the Ribaucour 
congruence of spheres. 

Algebraically, the Ribaucour transformation needs only a half of the data 
(necessary to build the congruence) of
the fundamental transformation. The data of the 
vectorial Ribaucour transformation consists of the solution 
$\boldsymbol{Y}^*_i:\mathbb{Z}^N\to\mathbb{V}^*$, 
of the linear system \eqref{eq:lin-H}. Then, because of the circularity
constraint, $\boldsymbol{Y}_i:\mathbb{Z}^N\to\mathbb{V}$ given by
\begin{equation*}
\boldsymbol{Y}_i = \left( 
\boldsymbol{\Omega}(\boldsymbol{X},\boldsymbol{Y}^*) +
T_i \boldsymbol{\Omega}(\boldsymbol{X},\boldsymbol{Y}^*) \right)^T
\bX_i, 
\end{equation*}
is a solution of the linear system \eqref{eq:lin-X}, and the constraints
\begin{gather*}
\boldsymbol{\Omega}(\boldsymbol{Y},H) +
\boldsymbol{\Omega}(X^\circ,\boldsymbol{Y}^*)^T=
2 \boldsymbol{\Omega}(\boldsymbol{X},\boldsymbol{Y}^*)^T
\bx, \\
\boldsymbol{\Omega}(\boldsymbol{Y},\boldsymbol{Y}^*) +
\boldsymbol{\Omega}(\boldsymbol{Y},\boldsymbol{Y}^*)^T = 
2 \boldsymbol{\Omega}(\boldsymbol{X},\boldsymbol{Y}^*)^T
\boldsymbol{\Omega}(\boldsymbol{X},\boldsymbol{Y}^*),
\end{gather*}
are admissible.

We remark that the above constraints have a simple geometric meaning when one
considers the circular lattices in $\mathbb{E}^M$  
as the stereographic projections of
quadrilateral lattices in the M\"obius sphere $S^M$; i.e., as a special case of
quadrilateral lattices subjected to quadratic constraints. 

\subsubsection{The symmetric lattice}
Given a quadrilateral lattice $\bx$ with rotation coefficients $Q_{ij}$ and
potentials $\rho_i$ given by \eqref{eq:rho-constr}, then the functions 
$\tilde{Q}_{ij}$, defined by equation
\begin{equation*}
\rho_j T_j \tilde{Q}_{ij} = \rho_i T_i Q_{ji}, \qquad i\ne j,
\end{equation*}
and called, because of their geometric interpretation, the backward rotation
coefficients, satisfy the Darboux system \eqref{eq:MQL-Q} as well. 
A quadrilateral
lattice is called \emph{symmetric} if its forward rotation coefficients 
$Q_{ij}$
are also its backward rotation coefficients. 
Again the constraint is compatible
with the geometric integrability scheme, i.e., it propagates in the construction
of the lattice. 
One can show that a quadrilateral lattice is symmetric if and only if its
rotation coefficients satisfy the following trilinear constraint
\begin{equation*}\label{eq:symm-cons-Q}
(T_iQ_{ji})(T_jQ_{kj})(T_kQ_{ik})=(T_jQ_{ij})(T_iQ_{ki})(T_kQ_{jk}),
\qquad i, j, k \quad \text{distinct}.
\end{equation*}

To obtain the corresponding reduction of the fundamental 
transformation we again
need only half of the data. Given a solution 
$\boldsymbol{Y}^*_i:\mathbb{Z}^N\to\mathbb{V}^*$, 
of the linear system \eqref{eq:lin-H}, then, because of the symmetric
constraint, $\boldsymbol{Y}_i:\mathbb{Z}^N\to\mathbb{V}$, defined by
\begin{equation*}
\boldsymbol{Y}_i = \rho_i(T_i\boldsymbol{Y}^*)^T ,
\end{equation*}
is the solution of the linear system \eqref{eq:lin-X}; notice that, equivalently,
we could start from $\boldsymbol{Y}_i$. The constraint
\begin{equation*}
\boldsymbol{\Omega}(\boldsymbol{Y},\boldsymbol{Y}^*) =
\boldsymbol{\Omega}(\boldsymbol{Y},\boldsymbol{Y}^*)^T ,
\end{equation*}
is then
admissible and gives a new
symmetric lattice. 

There are other multidimensional reductions of the quadrilateral lattice 
like,
for example, the $D$-invariant and Egorov lattices or discrete
versions of immersions of spaces of constant negative curvature. We remark
that the transformations and reductions discussed above have also a clear
interpretation on the level of the analytic methods.

\section{Integrable discrete surfaces}
\label{sec:surf-curv}
In this Section we present some distinguished examples of discrete integrable
surfaces. Notice that, although the geometric integrability scheme is 
meaningless for
$N=2$, however it can be applied indirectly,
by considering the discrete surfaces, together with their transformations,
as sub-lattices of multidimensional lattices. 

We remark  also that one can consider integrable
evolutions of discrete curves, which give equations of the Ablowitz--Ladik
hierarchy, and the corresponding integrable spin chains.
\subsection{Discrete isothermic nets}
\label{sec:isoth}
\begin{figure}
\begin{center}
\epsffile{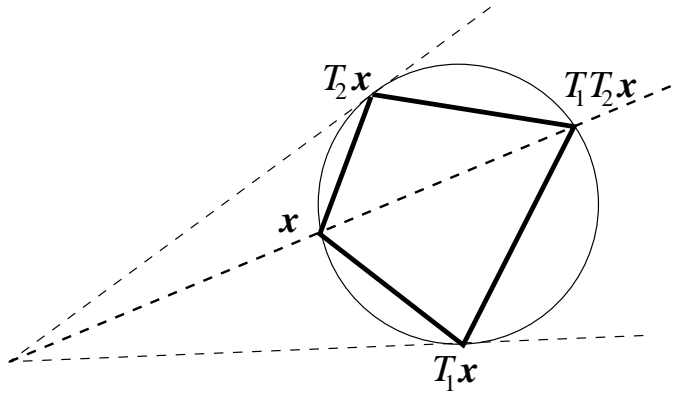}
\end{center}
\caption{Elementary quadrilaterals of the isothermic lattice}
\label{fig:harm-quad}
\end{figure}

An \emph{isothermic lattice} is 
a two dimensional circular lattice $\bx:\mathbb{Z}^2\to\mathbb{E}^M$ 
with harmonic quadrilaterals,; i.e., given $\bx$, $T_1\bx$ and $T_2\bx$, then
the point $T_1T_2\bx$ is the intersection of the circle (passing
through $\bx$, $T_1\bx$ and $T_2\bx$) and the line passing through 
$\bx$ and the
meeting point of the tangents to the circle at $T_1\bx$ and $T_2\bx$
(see Figure~\ref{fig:harm-quad}).
Therefore, given two discrete curves intersecting in the common vertex 
$\bx_0$, the unique isothermic lattice
can be found using the above "ruler and compass" construction.

Algebraically the reduction looks as follows. Any oriented plane in 
$\mathbb{E}^M$ can
be identified with the complex plane $\mathbb{C}$. Given any four complex 
ponts $z_1$,
$z_2$, $z_3$ and $z_4$, their complex cross-ratio is defined by
\begin{equation*}
q(z_1,z_2,z_3 ,z_4) = \frac{(z_1 - z_2)(z_3-z_4)}{(z_2 - z_3)(z_4 - z_1)}.
\end{equation*}
One can show that the cross-ratio is real if and only if the four points are
cocircular or collinear. In particular, a harmonic
quadrilateral with vertices numbered
anti-clockwise has cross-ratio equal to $-1$. Therefore, abusing the notation
(it can be formalized using Clifford algebras), the isothermic lattice is defined by
the condition 
\begin{equation*}
q(\bx,T_1\bx,T_1T_2\bx,T_2\bx) = -1.
\end{equation*}
We remark that the definition of isothermic lattices can be slightly generalized
allowing for the above cross-ratio to be a ratio of two real functions of single
discrete variables.

The restriction of the Ribaucour transformation to the
class of isothermic lattices,
named after Darboux who constructed it for isothermic surfaces,
has as its data a real parameter $\lambda$ and the starting point 
$\mathcal{D}(\bx_0)$, and can be described as follows. 
Given the elementary quadrilateral
$\{ \bx, T_1\bx, T_2\bx, T_1T_2\bx \}$ of the isothermic lattice, and
given the point $\mathcal{D}(\bx)$, then the 
points $\mathcal{D}(T_1\bx)$ and $\mathcal{D}(T_2\bx)$ 
belong to the corresponding planes and
are constructed from equations
\begin{align*}
q(\bx,\mathcal{D}(\bx),\mathcal{D}(T_1\bx),T_1\bx) & = \; \;\lambda, \\
q(\bx,\mathcal{D}(\bx),\mathcal{D}(T_2\bx),T_2\bx) & = - \lambda.
\end{align*}
It turns out that the point $\mathcal{D}(T_1 T_2\bx)$, constructed 
by the application of the geometric integrability scheme, is such that the
quadrilateral
$\{ \mathcal{D}(\bx), \mathcal{D}(T_1\bx), \mathcal{D}(T_2\bx),
\mathcal{D}(T_1 T_2\bx)\}$ is harmonic. Moreover,
the construction of the Darboux transformation is compatible; i.e.,
the new side quadrilaterals have the correct cross-ratios $\lambda$ and
$-\lambda$.

There are various integrable reductions of the isothermic lattice, for
example the constant mean curvature lattice and the minimal lattice. 
\subsection{Asymptotic lattices and their reductions}
\begin{figure}
\begin{center}
\epsffile{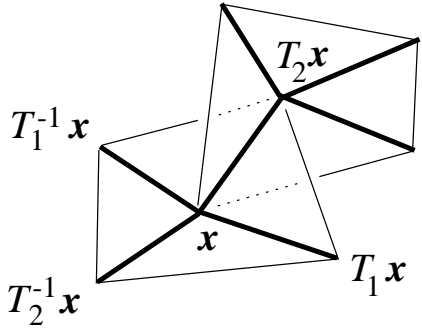}
\end{center}
\caption{Asymptotic lattices}
\label{fig:izot-cong}
\end{figure}

An \emph{asymptotic lattice} is a mapping $\bx:\mathbb{Z}^2\to\mathbb{R}^3$ such that any
point $\bx$ of the lattice is coplanar with its four nearest neighbours
$T_1\bx$, $T_2\bx$, $T_1^{-1}\bx$, $T_2^{-1}\bx$ (see
Figure~\ref{fig:izot-cong}). Such a plane is called the
tangent plane of the asymptotic lattice in the point $\bx$. 

It can be shown that any
asymptotic lattice $\bx$ 
can be recovered from its suitably rescaled 
normal (to the tangent plane) 
field $\boldsymbol{N}:\mathbb{Z}^2\to\mathbb{R}^3$ via the discrete 
analogue of the Lelieuvre formulas
\begin{equation}
\label{eq:Lelieuvre}
\Delta_1\bx = (T_1\boldsymbol{N})\times \boldsymbol{N},\qquad
\Delta_2\bx =  \boldsymbol{N}\times(T_2\boldsymbol{N}).
\end{equation}
By the compatibilty of the Lelieuvre formulas, the normal field $\boldsymbol{N}$
satisfies the discrete Moutard equation
\begin{equation} \label{eq:Moutard}
T_1 T_2 \boldsymbol{N} + \boldsymbol{N} =
 F ( T_1\boldsymbol{N} + T_2\boldsymbol{N}),
\end{equation}
for some potential $F:\mathbb{Z}^2\to\mathbb{R}$.

Given a scalar solution $\theta$ of the Moutard equation \eqref{eq:Moutard}, a
new solution $\mathcal{M}(\boldsymbol{N})$ of the Moutard equation, with the new
potential
\begin{equation*}
\mathcal{M}(F) = \frac{(T_1\theta)(T_2\theta)}{(T_1 T_2\theta)\theta} F,
\end{equation*}
can be found via the Moutard transformation equations
\begin{align}
\label{eq:Moutard-tr1}
\mathcal{M}(T_1 \boldsymbol{N})\mp \boldsymbol{N}& =
\frac{\theta}{T_1\theta}(\mathcal{M}(\boldsymbol{N}) \mp T_1\boldsymbol{N}), \\
\label{eq:Moutard-tr2}
\mathcal{M}(T_2 \boldsymbol{N})\pm \boldsymbol{N}& =
\frac{\theta}{T_2\theta}(\mathcal{M}(\boldsymbol{N}) \pm T_2\boldsymbol{N}).
\end{align}
Now, via the Lelieuvre formulas \eqref{eq:Lelieuvre},
one can construct a new asymptotic lattice 
$\mathcal{M}(\bx)=\bx \pm \mathcal{M}(\boldsymbol{N}) \times
\boldsymbol{N}$.
The lines connecting corresponding points of the asymptotic lattices $\bx$ and 
$\mathcal{M}(\bx)$ are tangent to both lattices. Such a $\mathbb{Z}^2$-family of lines
in $\mathbb{R}^3$ is called Weingarten (or $W$ for short) congruence. Notice that this
is not a congruence as considered in Section~\ref{sec:congr-fund}.

Various integrable reductions of asymptotic lattices are known in the
literature: pseudospherical lattices, 
asymptotic Bianchi lattices and
isothermally-asymptotic (or Fubini--Ragazzi) lattices, 
discrete (proper and improper) affine spheres.

Formally, the Moutard transformation is a reduction of the (projective
version of the) fundamental transformation for the Moutard reduction of the
Laplace equation. However the geometric relation between asymptotic lattices
and quadrilateral lattices is more subtle and 
the geometric scenery of this connection is the line geometry of
Pl\"{u}cker. Straight lines in $\mathbb{R}^3\subset\mathbb{P}^3$ 
are considered there as points of the so
called Pl\"{u}cker quadric $\mathcal{Q}_P\subset\mathbb{P}^5$.
A discrete asymptotic net in $\mathbb{P}^3$, viewed as the envelope of its tangent 
planes, corresponds to a congruence of isotropic lines
in $\mathcal{Q}_P$, whose focal lattices represent the asymptotic directions.
The discrete $W$-congruences are represented by
two dimensional quadrilateral lattices in the Pl\"{u}cker quadric.
\subsection{The Koenigs lattice}
\begin{figure}
\begin{center}
\epsffile{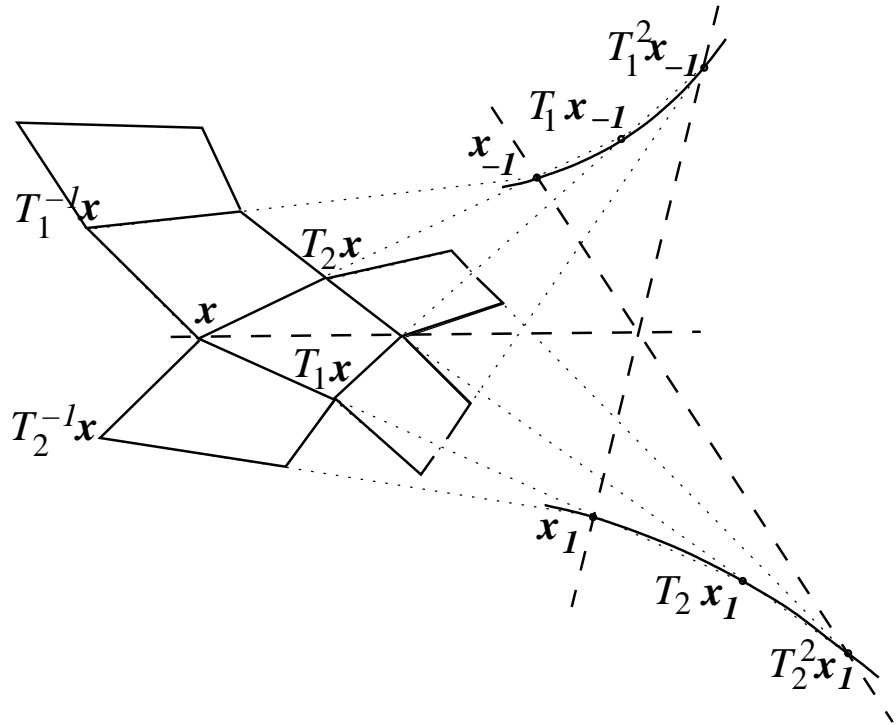}
\end{center}
\caption{The Koenigs lattice}
\label{fig:Koenigs}
\end{figure}

A two dimensional quadrilateral lattice $\bx:\mathbb{Z}^2\to\mathbb{P}^M$ is called a
\emph{Koenigs lattice} if, for every point $\bx$ of the lattice, 
the six points
$\bx_{\pm 1}$, $T_i\bx_{\pm 1}$, $T_i^2\bx_{\pm 1}$, $i=1,2$ of its Laplace
transforms (see Section~\ref{sec:Lapl-Hir}) belong to a conic (see
Figure~\ref{fig:Koenigs}).
The nonlinear constraint in definition of the Koenigs lattice can be
linearized, with the help of the Pascal "mystic hexagon"
theorem, to the form that the line passing through $\bx$ and $T_1T_2\bx$,
the line passing through $\bx_1$ and $T_1^2\bx_{-1}$, and
the line passing through $\bx_{-1}$ and $T_2^2\bx_1$ intersect in a point.  

Algebraically, the geometric Koenigs lattice condition means that the
Laplace equation of the lattice in homogeneous coordinates
$\boldsymbol{\mathrm{x}}:\mathbb{Z}^2\to\mathbb{R}^{M+1}_*$ can be gauged into the form
\begin{equation} \label{eq:Koenigs}
T_1 T_2\boldsymbol{\mathrm{x}} + \boldsymbol{\mathrm{x}} = 
T_1(F\boldsymbol{\mathrm{x}}) + T_2 (F\boldsymbol{\mathrm{x}}).
\end{equation}
It turns out that, if
$\boldsymbol{N}$ is a solution of the Moutard equation~\eqref{eq:Moutard}, 
then $\boldsymbol{\mathrm{x}}=T_1 \boldsymbol{N} + T_2 \boldsymbol{N}$ satisfies the 
Koenigs lattice equation.
Therefore, the algebraic theory of the discrete Koenigs lattice
equation~\eqref{eq:Koenigs}, its (Koenigs) transformation and the permutability
of the superpositions of such transformations is based on the corresponding
theory for the Moutard equation~\eqref{eq:Moutard}.

Geometrically, the Koenigs lattices 
are selected from the quadrilateral lattices as follows. 
Given a two dimensional quadrilateral
lattice $\bx:\mathbb{Z}^2\to\mathbb{P}^M$ and given a congruence 
$\mathfrak{l}$ with 
lines passing through the corresponding points of the lattice. Denote by
$\boldsymbol{y}_i = T_i^{-1}\mathfrak{l} \cap \mathfrak{l}$, $i=1,2$, 
points of the
focal lattices of the congruence. For every line $\mathfrak{l}$, denote by
$\imath$ the unique projective involution exchanging $\boldsymbol{y}_i$ with 
$T_i\boldsymbol{y}_i$. If, for every congruence $\mathfrak{l}$, 
the lattice $\mathcal{K}(\bx):\mathbb{Z}^2\to\mathbb{P}^M$,
with points $\mathcal{K}(\bx)=\imath(\bx)$, is a
quadrilateral lattice, then the lattice $\bx$ is a Koenigs 
lattice. The above construction
gives also the corresponding reduction of the fundamental transformation.

A distinguished reduction of the Koenigs lattice is the quadrilateral Bianchi
lattice. The natural continuous limit of the corresponding
equation is equivalent to the Bianchi (or hyperbolic Ernst)
system describing the interaction of planar gravitational waves. 

\subsection{Discrete two dimensional
Schr\"odinger equation}
In the previous secions we have discussed examples of integrable discrete
geometries described by equations of hyperbolic type. Below we present some
results associated with the elliptic case; it is remarkable that the QL provides
a way to connect these two subjects.

Consider a solution $\boldsymbol{N}: \mathbb{Z}^2\to\mathbb{R}^3$ of
the general self-adjoint 5-point scheme on the star of the 
$\mathbb{Z}^2$ lattice
\begin{equation} \label{eq:g-5p-s-a}
aT_1 \boldsymbol{N} +T^{-1}_1(a\boldsymbol{N})+
bT_2\boldsymbol{N}+T^{-1}_2(b\boldsymbol{N}) - c\boldsymbol{N} = 0,
\end{equation}
then the lattice $\boldsymbol{x}: \mathbb{Z}^2\to\mathbb{R}^3$ obtained 
by the Lelieuvre type formulas
\begin{equation}
\label{Lelieuvre2}
\Delta_1\boldsymbol{x}=-(T_2^{-1}b)\boldsymbol{N} \times T_2^{-1}\boldsymbol{N},
\qquad 
\Delta_2\boldsymbol{x}=(T_1^{-1}a)\boldsymbol{N} \times T_1^{-1}\boldsymbol{N},
\end{equation}
is a quadrilateral lattice having $\boldsymbol{N}$ as normal (to the planes of
elementary quadrilaterals) vector field.

The following gauge-equivalent form of equation \eqref{eq:g-5p-s-a}
\begin{equation}
\frac{\Gamma}{T_1\Gamma}T_1\psi+T^{-1}_1(\frac{\Gamma}{T_1\Gamma}\psi)+
\frac{\Gamma}{T_2\Gamma}T_2\psi+T^{-1}_2(\frac{\Gamma}{T_2\Gamma}\psi)-q\psi=0,
\end{equation}
an integrable discretization of the Schr\"odinger equation 
\begin{equation*}
\frac{\partial^2\psi}{\partial x_1^2}+
\frac{\partial^2\psi}{\partial x_2^2}-Q\psi=0, 
\end{equation*}
is also the Lax operator associated with an integrable generalization 
of the Toda law to the square lattice.

The 5-point scheme \eqref{eq:g-5p-s-a} is also a distinguished illustrative
example of the sub-lattice theory. Indeed it can be obtained restricting to the
even sub-lattice $\mathbb{Z}^2_e$  
the discrete Cauchy--Riemann equations
\begin{equation} \label{eq:d-C-R}
T_1T_2\phi - \phi = iG (T_1\phi-T_2\phi).
\end{equation}
Because of the equivalence (on the discrete level!) between 
equation \eqref{eq:d-C-R} and
the discrete Moutard equation \eqref{eq:Moutard}, the 5-point scheme 
\eqref{eq:g-5p-s-a} inherits integrability 
properties (Darboux-type transformations, superposition formulas, 
analytic methods of solution) from the 
corresponding (and simpler) integrability properties of the discrete 
Moutard equation.

\section*{Further Reading}
Akhmetshin, A.~A., Krichever, I.~M.  and Volvovski,  Y.~S. (1999), 
Discrete analogs of the {D}arboux--{E}goroff metrics, 
{\it Proc. Steklov Inst. Math.}  225, 16-39.

\bigskip\noindent
Bia{\l}ecki, M. and Doliwa, A (2004), Algebro-geometric solution of the discrete
KP equation over a finite field out of a hyperelliptic curve, {\it Comm. Math.
Phys.}  253, 157-170.

\bigskip\noindent
Bobenko,~A.I. and Seiler,~R. (eds.) (1999), {\it Discrete Integrable 
Geometry and Physics}, Clarendon Press, Oxford.

\bigskip\noindent
Bobenko,~A.I. (2004), Discrete differential geometry. Integrability as
consistency, In Grammaticos, B., Kosmann-Schwarzbach Y. and Tamizhmani T. (eds.)
{\it Discrete Integrable Systems}, pp. 85-110, Springer, Berlin

\bigskip\noindent
Bogdanov, L. V. and Konopelchenko, B. G. (1995),
Lattice and q-difference Darboux--Zakharov--Manakov systems via
$\bar{\partial}$ method,
{\it J. Phys.} A28, L173-L178.

\bigskip\noindent
Cie\'{s}li\'{n}ski, J. (1997), The spectral interpretation of $N$-spaces of
constant negative curvature immersed in 
$\mathbb{R}^{2N-1}$, {\it Phys. Lett. } A236,
425--430.

\bigskip\noindent
Doliwa, A., Ma\~nas, M., Mart\'{\i}nez Alonso, L.,
Medina, E. and Santini, P. M. (1999),
Charged Free Fermions,
Vertex Operators and Transformation Theory of Conjugate Nets,
{\it J. Phys.} A32, 1197--1216.

\bigskip\noindent
Doliwa,~A., Santini,~P.M. and Ma{\~n}as,~M. (2000),
Transformations of Quadrilateral Lattices, {\it J. Math. Phys.} 41, 
944--990.

\bigskip\noindent
Doliwa,~A. and Santini,~P.M. (2000), The symmetric, $D$-invariant and Egorov
reductions of the quadrilateral latice, {\it J.~Geom. Phys.} 36, 60--102.

\bigskip\noindent
Doliwa,~A., Nieszporski,~M and Santini,~P.M. (2001), 
Asymptotic lattices 
and their integrable reductions I: the Bianchi and the Fubini-Ragazzi 
lattices, {\it J. Phys.} A34, 10423--10439.

\bigskip\noindent
Doliwa,~A., Nieszporski,~M and Santini,~P.M. (2004), 
Geometric discretization of the Bianchi  
system, {\it J. Geom. Phys.} 52, 217--240.

\bigskip\noindent
Doliwa, A., Grinevich, P. G., Nieszporski, M. and Santini, P. M. (2004), 
Integrable lattices and their sublattices: from the discrete Moutard 
(discrete Cauchy-Riemann) 4-point equation to the self-adjoint 5-point scheme, 
{\tt nlin.SI/0410046}.

\bigskip\noindent
Ma\~{n}as, M. (2001), Fundamental transformation for quadrilateral lattices:
first potentials and $\tau$-functions, symmetric and pseudo-Egorov reductions,
{\it J.~Phys.} A34, 10413--10421.

\bigskip\noindent
Matsuura, N. and Urakawa, H. (2003), Discrete improper affine spheres, {\it
J.~Geom. Phys.} 45, 164--183.

\bigskip\noindent
Rogers,~C. and Schief,~W.~K. (2002), {\it B\"{a}cklund and Darboux
Transformations. Geometry and Modern Applications in Soliton Theory},
University Press, Cambridge.

\bigskip\noindent
Schief,~W.~K. (2003), On the unification of classical and novel integrable
surfaces: II. Difference geometry, {\it Proc. R. Soc. London} A459, 373-391. 

\bigskip\noindent
Schief,~W.~K. (2003), Lattice Geometry of the Discrete Darboux, KP, BKP and CKP
Equations. Menelaus' and Carnot's Theorems, {\it J.~Nonlin. Math. Phys.} 10,
Supplement 2, 194--208. 

\newpage

\listoffigures


\section*{See also}
B\"{a}cklund transformations\\
$\bar\partial$ approach to integrable systems\\
Discrete integrable systems\\
Integrable systems and algebraic geometry\\
KP equations and geometry\\
Nonlinear Schr\"{o}dinger equations\\
Sine-Gordon equation\\
Toda lattices


\section*{Keywords}

Integrable discrete geometry\\
Quadrilateral lattice\\
Line congruences\\
Fundamental transformation\\
Circular lattice\\
Symmetric lattice\\
Isothermic lattices\\
Asymptotic lattices\\
Koenigs lattice\\
Discrete 2D Schr\"odinger equation

\end{document}